# Reconfigurable Generation of Terahertz Spiral Vector Beams Based on All-Dielectric Cascaded Metasurfaces

HONGJUN CHEN,[1,2] YUXIN TANG,[1,2] JIANMING LI,[1,2] JITAO LI,[3] LI LUO,[1,2] TINGTING TANG,[1,2] JIE LI,[1,2,*] JIANQUAN YAO[4]

[1] Sichuan Province Key Laboratory of Optoelectronic Sensor Devices and Systems, College of Optoelectronic Engineering (Chengdu IC Valley Industrial College), Chengdu University of Information Technology, Chengdu 610225, China.

[2] Sichuan Meteorological Optoelectronic Sensor Technology and Application Engineering Research Center, Chengdu University of Information Technology, Chengdu 610225, China.

[3] School of Science, Southwest Petroleum University, Chengdu 610500, China.

[4] School of Precision Instruments and Opto-Electronics Engineering, Tianjin University, Tianjin 300072, China.

## Abstract

The reconfigurable generation of complex structured terahertz (THz) fields, and vector beams in particular, remains a significant challenge. Here, we propose a novel all-dielectric cascaded-metasurface scheme for the reconfigurable generation of THz spiral vector beams. By combining an isotropic phase-compensation layer with an anisotropic circular-polarization-sensitive layer, together with a new phase design tailored to spiral beams, we achieve independent control of the left- and right-handed circularly polarized (LCP and RCP) components through the relative rotation angle between the two layers. Under linearly polarized THz wave illumination, THz vector beams with switchable spiral numbers of 1-3 and reversible circular-polarization handedness are demonstrated at frequencies around 0.9 THz. Moreover, owing to the longitudinal evolution characteristic of spiral beams, the three generated THz vector beams exhibit azimuthal rotations of 279.9°, 125.7° and 75.5°, respectively, along the propagation direction. Such complex THz vector beams, which combine reconfigurability with longitudinal evolution, are expected to find applications in emerging fields such as next-generation wireless communications and THz imaging. The proposed generation scheme also offers a useful reference for the reconfigurable generation of high-dimensional vector beams.

Keyword： Terahertz wave; structured optical field; vector beam; Moiré phase engineering， metasurface

## Introduction

Structured light fields are a class of tailor-made optical fields with distinctive spatiotemporal evolution characteristics. During propagation and in their interaction with matter, they exhibit a variety of novel properties, such as photonic spin-orbit coupling, topological singularities, mechanical effects and modal orthogonality, and have therefore been widely applied in microparticle manipulation, biomedical imaging, optical communication and sensing, and laser processing. Structuring of an optical field can be achieved by introducing phase and polarization singularities into a beam. As research advances and applications expand, higher demands have been placed on the design dimensions and degrees of freedom available for structured light fields [1-3]. Consequently, key techniques for generating and manipulating beams in which multiple electromagnetic parameters vary simultaneously across multiple spatiotemporal dimensions, as well as for the real-time reconfigurable generation of differently structured beams, are of increasing value and significance. Structured light fields in the terahertz (THz) regime also offer distinctive application prospects. For example, in next-generation (6G) wireless communications, the orbital angular momentum (OAM) of THz beams provides an additional degree of freedom that can further increase data rates and channel capacity [4]. In the field of novel electron acceleration, high-power THz vector beams open new possibilities for electron guns, accelerators, and beam manipulation and diagnostics; THz-driven electron acceleration has thus become one of the most promising routes towards future high-quality, compact accelerators [5]. In imaging, structured light fields such as THz vortex beams, Bessel beams and Airy beams provide essential support for techniques including high-resolution, extended-depth-of-focus and high-contrast THz phase-contrast imaging and computational tomography [6]. The efficient generation of structured light fields in the THz regime is a prerequisite for advancing the broader application of THz technology, and the reconfigurable generation of diverse THz structured beams remains a critical issue to be resolved.

In conventional optical systems, the reconfigurable generation of structured light fields, and vector beams in particular, generally requires multiple steps and involves devices such as spatial light modulators, polarizers and wave plates, together with beam-splitting and beam-combining operations. Because the THz band has long lacked mature digital spatial modulation devices, early generation of THz structured light fields relied mainly on beam shaping of conventional beams using devices such as phase plates, cascaded wave plates and liquid crystals [7–10]. Such approaches can hardly control two or more electromagnetic parameters jointly, and the independent design of different field components remains difficult; they therefore still cannot meet the requirements for generating high-dimensional THz structured light fields. In recent years, advances in metasurface research have opened new opportunities for the generation of high-dimensional THz structured light fields. In 2017, single-layer metallic metasurfaces were employed for the multiplexing–demultiplexing of THz orbital angular momentum and for the generation of radially polarized Lorentz beams [11–14]. In 2020, Han et al. combined a nonlinear metasurface with an epsilon-near-zero material to directly excite THz vortex beams at the radiation source [15]. In 2021, Yao et al. used all-silicon metasurfaces to generate THz orbital-angular-momentum superpositions multiplexed in linear and circular polarizations, respectively [16–18]. Nevertheless, such single-layer metasurfaces still struggle to realize reconfigurable generation of THz structured beams. Bilayer moiré metasurfaces offer an effective solution to this problem. For example, in 2023, Chen et al. proposed a mechanically rotated cascaded-metasurface scheme and demonstrated dynamic control of THz Airy beams and focused Gaussian beams in the transverse and longitudinal dimensions, respectively [19]. Also in 2023, Wang et al. used a bilayer metasurface to flexibly manipulate the order of Bessel beams, thereby achieving wide-range modulation of the non-diffracting propagation length [20]. In 2025, Nan et al. further achieved simultaneous switching of the polarization and topological charge of THz Bessel beams [21]. In 2026, Xu et al. likewise employed a bilayer cascaded metasurface to realize continuous in-plane focal shifting with a broad tuning range and broadband compatibility [22]. However, these studies have focused mainly on the reconfigurable generation of scalar THz structured beams, and research on vector-beam generation remains markedly insufficient.

To address these issues, we propose a cascaded moiré metasurface device operating in the THz regime. A circular-polarization-multiplexing layer provides polarization decoupling, while a polarization-independent layer switches the overall phase of each channel; combined with a specifically designed phase profile for spiral beams, this configuration generates THz vector beams whose spiral number and polarization-state distribution are dynamically tunable. By adjusting the relative rotation angle $\Delta\beta$ between the two metasurfaces, the number of azimuthal spiral spots can be switched and the circular-polarization handedness reversed.

## Results and Discussion

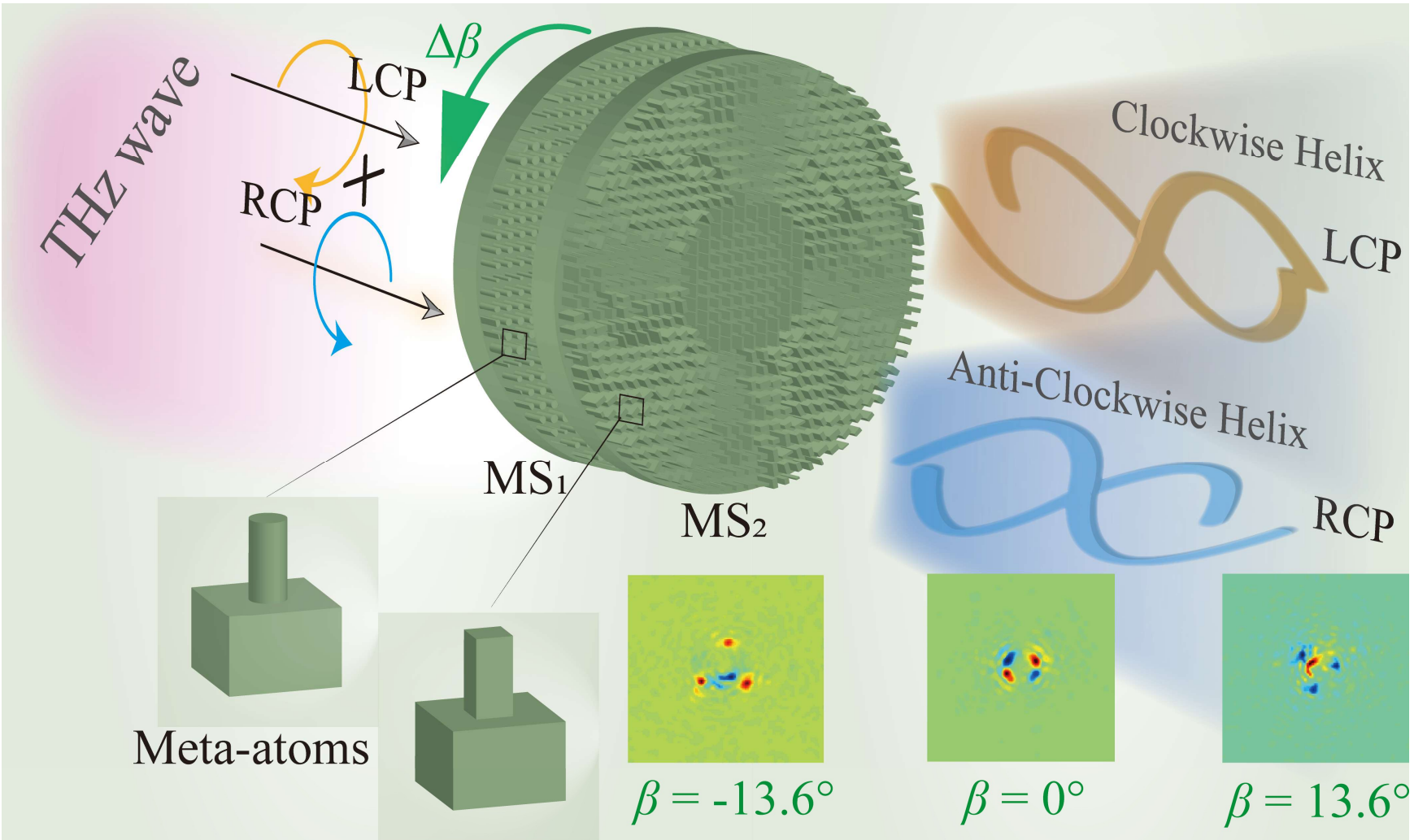

**Figure 1** Schematic of the reconfigurable generation of THz spiral vector beams based on a cascaded metasurface.

We propose a polarization-multiplexed all-dielectric moiré metasurface scheme, in which the lens comprises two coaxially aligned metasurfaces. The metasurface is designed to impart two distinct transfer functions, respectively to the orthogonal left- and right-hand circularly polarized (LCP and RCP) THz waves, so that illumination with a linearly polarized beam yields vector beams whose circular polarization is spatially non-uniformly distributed. As shown in Fig. 1, the designed cascaded metasurface consists of two all-dielectric layers: MS1 is an isotropic, polarization-insensitive layer, and MS2 is an anisotropic, circular-polarization-sensitive layer that provides phase modulation for the orthogonal circularly polarized waves. Through a tailored phase design, vector beams with conjugate spiral handedness and different spiral numbers are generated as the relative rotation angle $\Delta\beta$ is varied. At $\Delta\beta = 0°$, the LCP component forms a clockwise double-helix beam whereas the RCP component forms a counter-clockwise double-helix beam, and the Stokes parameter S3 at the focal plane exhibits two types of patches distributed in a uniform alternating pattern. At $\Delta\beta =$ 13.6°or $\Delta\beta = -13.6°$, the LCP component becomes a clockwise triple-helix or single-helix beam, respectively, while the RCP component becomes a counter-clockwise single-helix or triple-helix beam. Consequently, under linearly polarized THz illumination, switching $\Delta\beta$ from 13.6°to −13.6° reverses the circular-polarization handedness of the inner and outer helical beams.

To achieve reconfigurable generation of spiral THz vector beams, we construct a polarization-sensitive propagation function whose spiral order is tunable (the amplitude differences among individual meta-atoms are neglected, and the transmission amplitude is assumed to be unity). The first metasurface, MS1, of the moiré metasurface acts as a phase-compensation element, and the number of generated spiral beams is controlled by the rotation angle β. Because MS1 is composed of polarization-insensitive cylindrical pillars, wavefront manipulation is realized through the propagation-phase principle. The second metasurface, MS2, is sensitive to circular polarization and is responsible for independently controlling the left-hand circularly polarized (LCP) and right-hand circularly polarized (RCP) components. According to the spin-decoupled phase-design principle[23], a rotation-angle-dependent phase is imparted to the LCP and RCP components.

The required phase function of the spiral beam is composed of the two independent components shown below: a hyperbolic lens phase (Eq. 1.1) and a two-fold self-rotational phase (Eq. 1.2) [24, 25]:

$$\varphi_{f_0}(x,y)=\frac{2\pi}{\lambda}\cdot\left(f_0-\sqrt{x^2+y^2+f_0^{\,2}}\right) \tag{1.1}$$

where $\lambda$ denotes the target wavelength of the incident light, $f_0$ is the design focal length, and x and y are the Cartesian coordinates on the metasurface plane.
The two-fold rotational phase distribution ($\varphi_{rot}$) is given by [26 – 28]:

$$\varphi_{rot}(r,\theta)=\pm k\cdot round\left(a\cdot r^b\right)\cdot\left(\theta+\frac{\pi}{2}\right) \tag{1.2}$$

where $k$ is a control parameter that determines the number of lobes into which the pattern is divided, $r$ is the radial coordinate ($r=\sqrt{x^2+y^2}$), θ is the azimuthal angle ($\theta = \tan^{-1}(y/x)$), round() denotes the nearest-integer rounding function, $a$ adjusts the tail of the rotating beam, and $b$ modifies the overall phase map. The optimal values were finally chosen as $a$ = 0.65 and $b$ = 1, and the sign of $\varphi_{rot}$ determines the sense of the spiral rotation, i.e. clockwise or counter-clockwise [29].
To make the spiral number tunable through the phase shift introduced by rotating the metasurface through an angle $\beta$, it is sufficient to let $\beta$ modify $k$; the phase distribution designed for MS1 is therefore:

$$\varphi_{MS1}=\varphi_{\mathrm{rot}}(r,\theta)\cdot\theta \tag{1.3}$$

The phase distribution of MS2 is given by:

$$\begin{aligned}\varphi_{MS2-LCP}&=-\varphi_{\mathrm{rot}}(r,\theta)\cdot(\theta+\beta)-\varphi_{f_0}(x,y)\\ \varphi_{MS2-RCP}&=-\varphi_{\mathrm{rot}}(r,\theta)\cdot(\theta-\beta)-\varphi_{f_0}(x,y)\end{aligned} \tag{1.4}$$

Summing the LCP and RCP phases of MS1 and MS2 separately yields the total phase distributions for the LCP and RCP channels:

$$\varphi_{joint-LCP} = -\beta \cdot k \cdot round\left(a \cdot r^b\right) \cdot \left(\theta + \frac{\pi}{2}\right) - \varphi_{f_0}\left(x, y\right)$$
$$\varphi_{joint-RCP} = \beta \cdot k \cdot round\left(a \cdot r^b\right) \cdot \left(\theta + \frac{\pi}{2}\right) - \varphi_{f_0}\left(x, y\right) \quad (1.5)$$

where $\varphi_{\text{joint-LCP}}$ and $\varphi_{\text{joint-RCP}}$ are the total phase distributions corresponding to the LCP and RCP components, respectively, and $\beta$ is the relative rotation angle. This yields a phase distribution with a tunable spiral number, which is approximately given by:

$$N = \text{round}\left(\left|\beta \cdot k\right|\right) \quad (1.6)$$

In this work, $k$ is kept fixed at 2. To obtain spirals with numbers 1, 2 and 3, and for experimental feasibility, we choose equal rotation-angle intervals; accordingly:

$$\beta = \left\{\frac{11\pi}{66}, \frac{21\pi}{66}, \frac{31}{66}\right\} \quad (1.7)$$

In this work, the initial configuration of the metasurface is chosen to correspond to $N$ = 2, i.e. $\beta$ = 21π/66, giving $\Delta\beta$ = ±10π/66. Because the initial phase is set to $N$ = 2, the signs of $\beta$ associated with the LCP and RCP channels differ in Eq. (1.5), whereas the phase loaded by the moiré lens is the same for the same sense of rotation; we therefore obtain:

$$N_{LCP} = round\left(\left|k \cdot \left(-\frac{21\pi}{66} \mp \Delta\beta\right)\right|\right)$$
$$N_{RCP} = round\left(\left|k \cdot \left(\frac{21\pi}{66} \mp \Delta\beta\right)\right|\right) \quad (1.8)$$

This yields $N_{LCP}$ = {3, 2, 1} and $N_{RCP}$ = {1, 2, 3}. We therefore obtain a phase distribution for which, under the same rotation angle, the LCP and RCP components have different spiral numbers and opposite senses of spiral rotation.

To achieve co-polarized phase control with MS1 and independent cross-polarized control with MS2, we designed two metasurface structures, namely cylindrical and rectangular pillars, both operating at 1 THz. The two silicon pillars were simulated with the time-domain solver of CST Studio Suite 2019, applying periodic boundary conditions in the x and y directions and using a plane wave as the excitation. The silicon pillars have a period P = 140 μm, a height H = 200 μm and a substrate thickness h = 300 μm. As shown in Fig. 2e, the transmission amplitude and phase were calculated by varying the cylinder diameter R1. From the simulation results, eight meta-atoms were selected (their structural parameters are listed in Table 1), and their dimensions together with the corresponding transmission amplitudes and phases are plotted in Fig. 2e. It can be seen that the transmission amplitude remains above 0.6, while the phase varies continuously over 2π in steps of π/4. For the rectangular pillars, the corresponding transmission amplitude and phase were obtained by varying the length *L* and width *W*. The silicon rectangular pillars must not only satisfy the half-wave-plate condition (i.e. equal amplitudes and a phase difference of π between the two orthogonally polarized transmitted co-polarized components) so as to enable polarization control, but must also provide a dynamic phase that can evolve continuously over 0–2π, so that the required phase distribution can be obtained. As shown in Fig. 2f, after screening the simulation data, fifteen meta-atoms were selected; their transmission amplitude remains above 0.6, while the phase varies continuously over 2π in steps of π/8. The corresponding transmission amplitudes and phases are shown in Fig. 2(a–d). At an incident frequency of 1 THz, we constructed a database of transmission amplitudes and phase shifts ($T_{xx}$, $T_{yy}$, $\varphi_{xx}$, $\varphi_{yy}$), recording the corresponding values as the rectangular-pillar dimensions ($W$, $L$) were varied. Here, $T_{xx}$ ($T_{yy}$) and $\varphi_{xx}$ ($\varphi_{yy}$) denote, respectively, the transmission amplitude and phase of the transmitted x-polarized (y-polarized) component under incident x-polarized (y-polarized) waves.

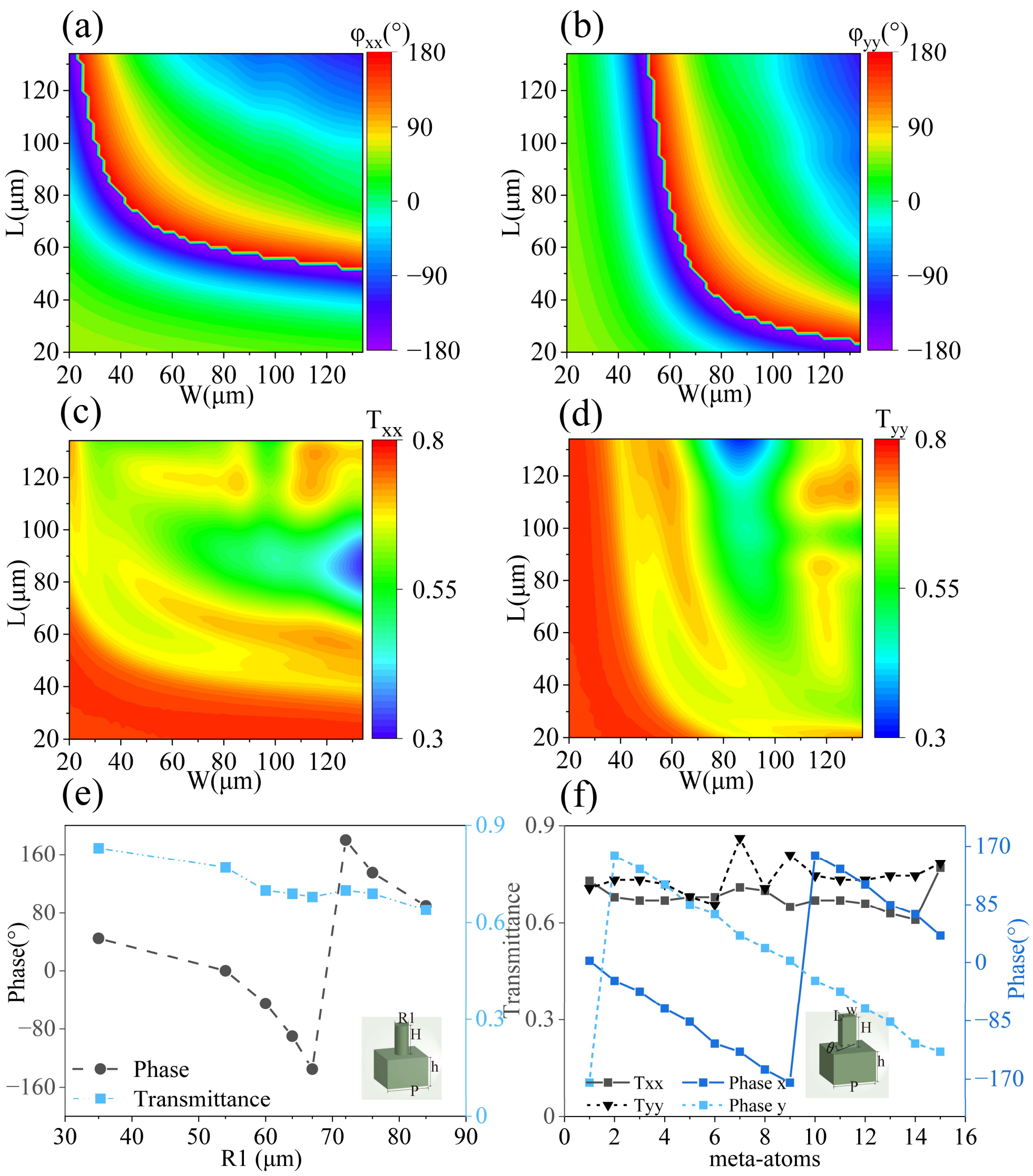

**Figure 2.** Sweep of the geometric parameters of the metasurface unit cells. (a, b) Transmission amplitude under x- and y-polarized incidence for meta-atoms of different sizes; (c, d) the corresponding phase shift; (e) amplitude and phase of the eight selected isotropic unit cells; (f) amplitude and phase of the sixteen selected anisotropic unit cells.

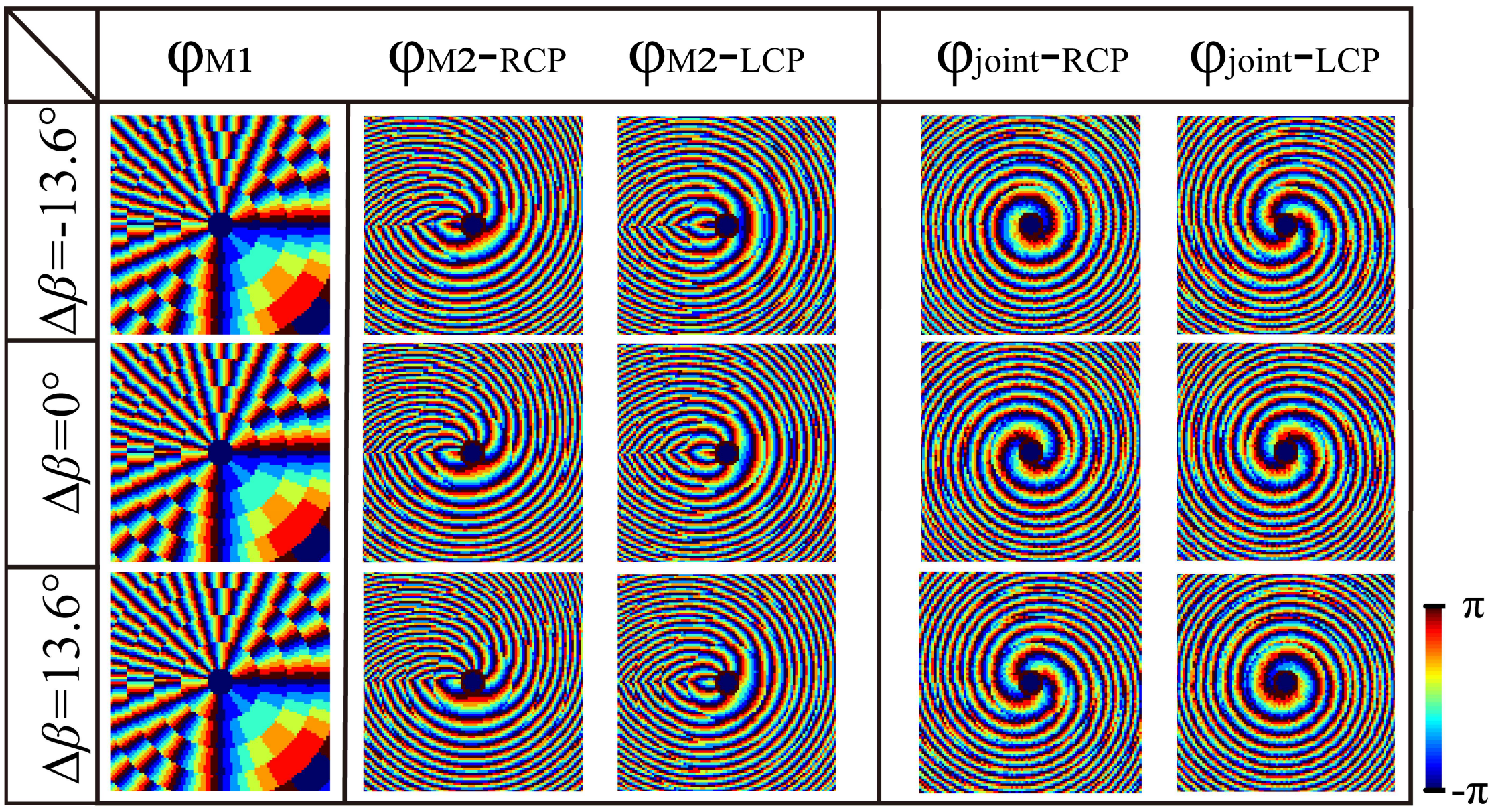

**Figure 3**. Design phases of the cascaded metasurfaces.

Based on the two selected meta-atom shapes, the unit cells of the two metasurfaces were arranged one by one to obtain the designed phase distributions of the two layers shown in Fig. 3. Since the initially chosen phase distribution corresponds to $k$ = 2, in the following description $\Delta\beta$ = 0° always refers to the initial relative rotation angle $\beta$ = 21π/66. Figure 3 shows the phase distributions of the moiré metasurface: the first to third columns are the phases of the two metasurfaces at the different rotation angles $\beta$, and the fourth and fifth columns are the total phase distributions corresponding to the different rotation angles $\beta$.

The designed two-layer cascaded metasurface was modelled in CST Studio Suite 2019 and simulated with the transient solver, with a designed focal length $f_0$ = 7 mm and an array size of [ 90×90 ].Figures 4 and 5 show the resulting field intensity (i.e. $|E|^2$) obtained under illumination by LCP and RCP terahertz waves, respectively, at a frequency of 0.9 THz and an incident field amplitude of 1 V/m. The time-domain pulse used in the simulation had a bandwidth of 0.6–1.2 THz; the best performance was obtained at 0.9 THz, and the performance at 1 THz was inferior to that at 0.9 THz. Analysis of the unit-cell library suggests that the probable cause is a deviation in the phase selection of some unit cells, which introduces an error between the overall phase of the element array and the design theory, thereby shifting both the optimum frequency and the optimum rotation angle. Similar simulation results were obtained over the range 0.85–1 THz, so we selected 0.9 THz and $|\Delta\beta| = 13.6°$ — the frequency and rotation angle giving the highest intensity — for presentation. Following the above analysis, Fig. 4 shows the modulation of 0.9 THz LCP incidence by the designed metasurface device. Specifically, for a relative rotation angle $\Delta\beta = 0°$, the output LCP wave is a double-helix beam rotating clockwise along the +z direction; for $\Delta\beta = -13.6°$, it is a clockwise triple-helix beam; and for $\Delta\beta = +13.6°$, it is a clockwise single-helix beam.

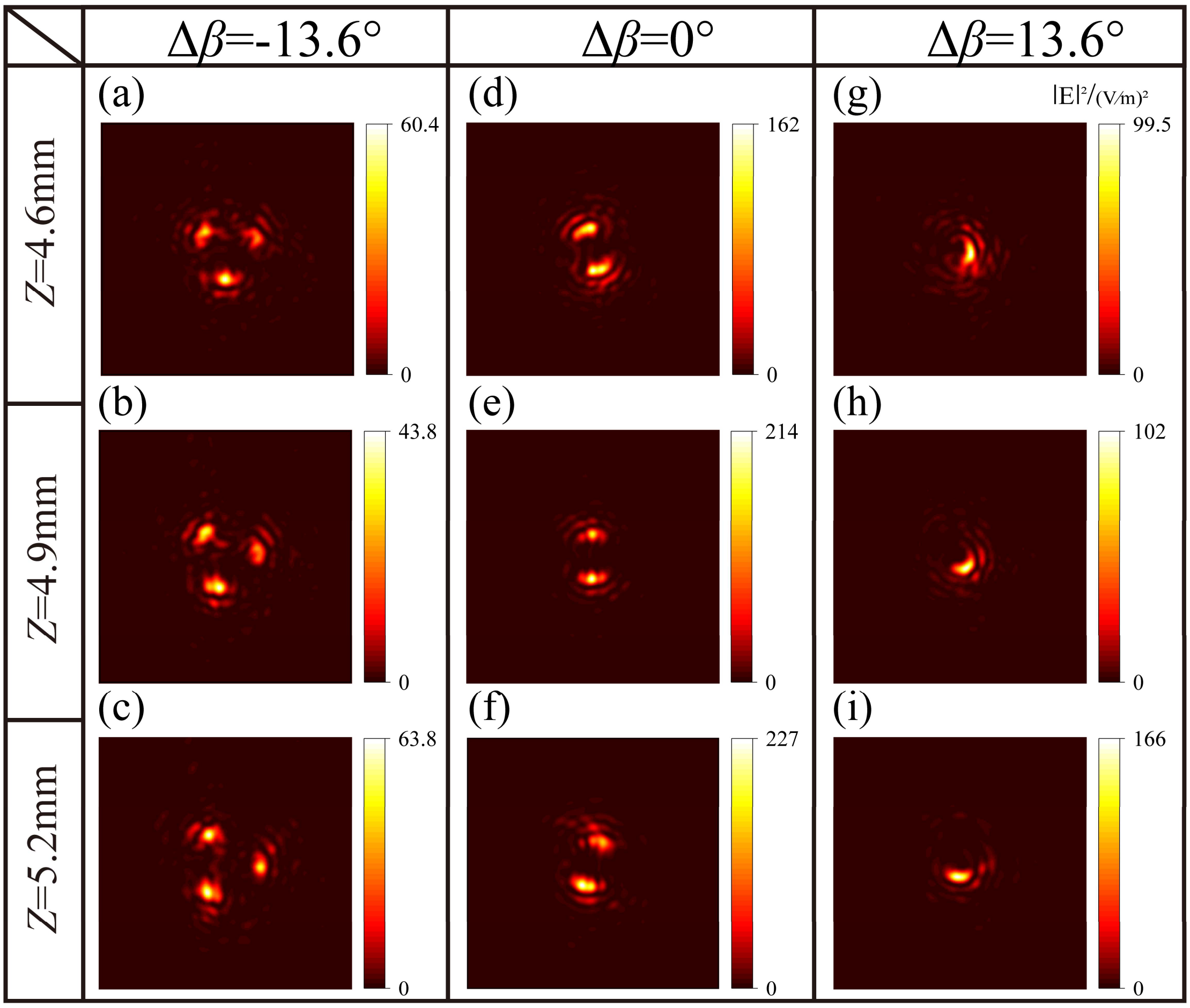

**Figure 4** Intensity distributions $|E|^2$ of the transmitted field under LCP incidence at different propagation distances (units: $V^2/m^2$). (a–c) Intensity distributions at the planes $Z = 4600$ μm, 4900 μm and 5200 μm for LCP incidence with $\Delta\beta = -13.6°$. (d–f) Intensity distributions at the same planes for LCP incidence with $\Delta\beta = 0°$. (g–i) Intensity distributions at the same planes for LCP incidence with $\Delta\beta = -13.6°$.

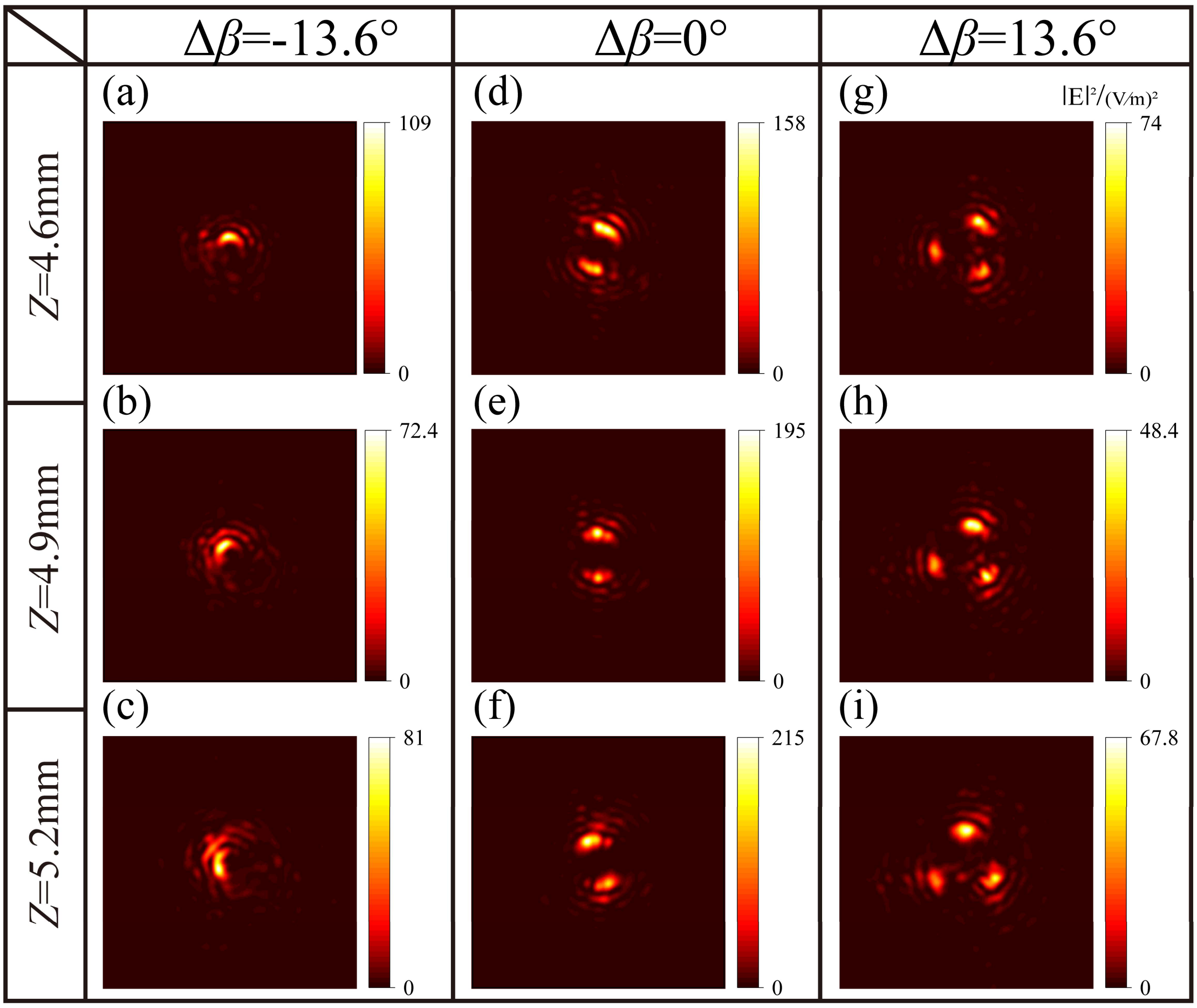

**Figure 5** Intensity distributions $|E|^2$ of the transmitted field under RCP incidence at different propagation distances (units: $V^2/m^2$).Intensity distributions at the planes $Z$ = 4600 μm, 4900 μm and 5200 μm for RCP incidence with $\Delta\beta = -13.6°$. (d–f) Intensity distributions at the same planes for RCP incidence with $\Delta\beta$ = 0°. (g–i) Intensity distributions at the same planes for RCP incidence with $\Delta\beta = -13.6°$.

Figure 5 shows the corresponding modulation of 0.9 THz RCP incidence; likewise, the helical beams output for RCP incidence rotate counter-clockwise along the +z direction. For $\Delta\beta$ = 0° the output is a double helix, for $\Delta\beta = -13.6°$ a single helix, and for $\Delta\beta = +13.6°$ a triple helix. A combined examination of Figs. 4 and 5 further reveals that the rotation rate of the helix varies with the number of helical beams; this will be analysed in detail below.

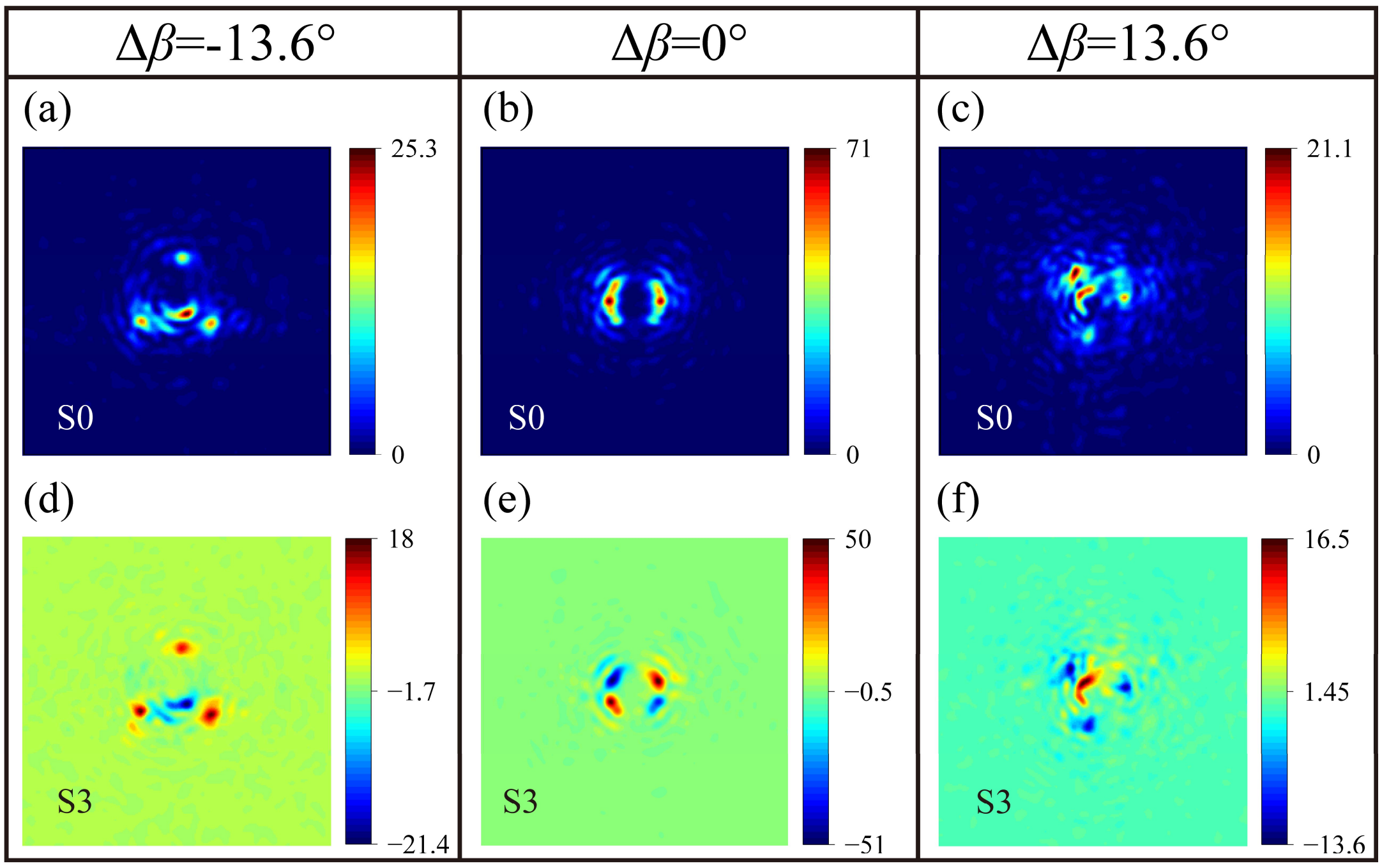

**Figure 6** Distributions of the Stokes parameters $S_0$ and $S_3$ in the field transmitted through the cascaded metasurface under linearly polarized incidence (a–c) Intensity distributions of the Stokes parameter $S_0$ (total intensity) at the planes $Z$ = 5.2 mm, 4.5 mm and 4.2 mm for different rotation angles $\Delta\beta$. (d–f) Distributions of the Stokes parameter $S_3$ (circular polarization component) at the same planes for different rotation angles $\Delta\beta$.

Figure 6 shows the calculated Stokes parameters $S_0$ and $S_3$ of the transmitted field at the planes $Z$ = 5.2 mm, 4.5 mm and 4.2 mm, for the designed cascaded metasurface illuminated by a linearly polarized plane terahertz wave with an amplitude of 1.4 V/m. It can be seen that, after the metasurface is rotated by different angles and illuminated with linearly polarized light, the transmitted beam still follows the same behaviour as when the two orthogonal circular polarizations are incident separately. From Fig. 6(d–f), for $\Delta\beta = 0°$ the $S_3$ distribution on the $Z$ plane consists of alternating LCP and RCP regions in the azimuthal direction. For $\Delta\beta = -13.6°$, the outer triple-helix beam is LCP while the inner single-helix beam is RCP. For $\Delta\beta = +13.6°$, the polarization distribution is exactly reversed with respect to that at $\Delta\beta = -13.6°$. We therefore obtain a scheme in which the polarization distribution can be reconfigured by rotating the cascaded metasurface, yielding reconfigurable terahertz helical vector beams.

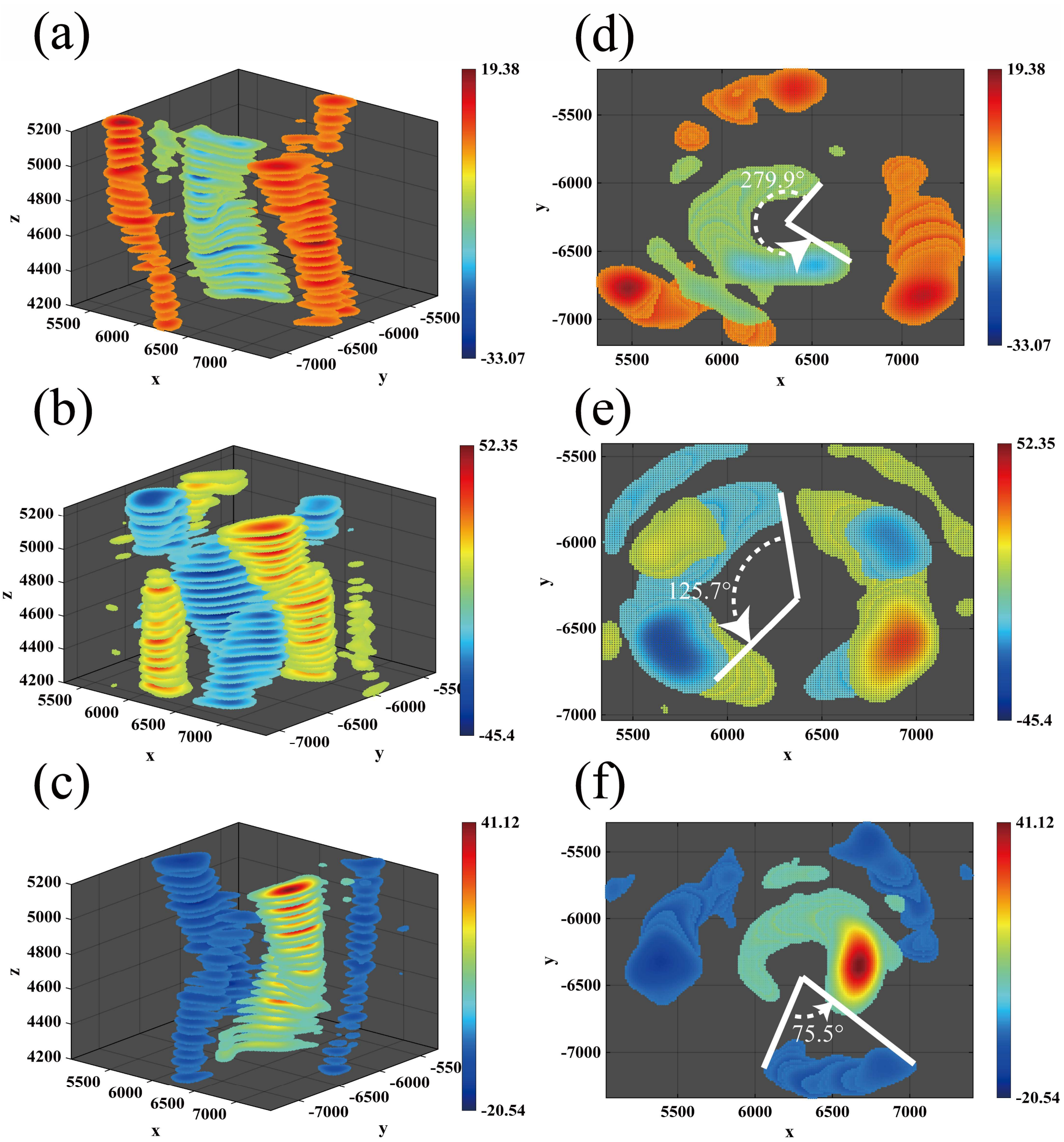

**Figure 7** Three-dimensional reconstruction of the $S_3$ parameter of the generated vector beams. (a) Three-dimensional view of $S_3$ for $\Delta\beta = -13.6°$. (b) Three-dimensional view of $S_3$ for $\Delta\beta = 0°$. (c) Three-dimensional view of $S_3$ for $\Delta\beta = +13.6°$. (d) Rotation angle of the $S_3$ distribution, viewed from above after the 3-D reconstruction, after a propagation of 1000 μm along $Z$, for $\Delta\beta = -13.6°$. (e) The same for $\Delta\beta = 0°$. (f) The same for $\Delta\beta = +13.6°$.

Figure 7 shows the three-dimensional view obtained by stacking the $S_3$ slices of the field data calculated on different $Z$ planes, spanning $Z = 4200$ μm to 5200 μm with a step of 50 μm. The three-dimensional display of the $S_3$ distribution allows the details of the helical beams under linearly polarized incidence — including their number and rotation sense — to be seen more intuitively. Figures 7(a)–(c) respectively show the $S_3$ distributions for $\Delta\beta = -13.6°$, 0° and +13.6°. From Fig. 7(a), at $\Delta\beta = -13.6°$ the LCP component forms a triple-helix beam rotating clockwise along +*Z*, whereas the RCP component forms a single-helix beam rotating counter-clockwise along +*Z*. These helix numbers are exactly reversed in Fig. 7(c) for the LCP and RCP components. The same behaviour is observed in Fig. 7(b) for both the LCP and RCP components. It is worth noting that within the range $Z$ = 4600–4800 μm the beam appears to disappear. The reason is that the double-helix beams of the two channels coincide in this region; because no phase difference was designed between the LCP and RCP components, their superposition there yields an ordinary linearly polarized field. The $S_0$ display in Fig. 6(b) appears to be consistent with this inference. A combined analysis of Figs. 7(a)–(f) resolves the issue left open in the discussion of Fig. 6, namely the rotation rate. Over z = 4200–5200 μm, the rotation is fastest for one helix (≈279.9°), intermediate for two helices (≈125.7°) and slowest for three helices (≈75.5°).

## Conclusion

In summary, a cascaded metasurface based on moiré phase engineering was designed, and the characteristics of its transmitted field under linearly polarized terahertz incidence were analyzed. The results show that the proposed scheme enables reconfigurable generation of helical vector beams in the terahertz band. By changing the relative rotation angle of the cascaded metasurface, the number of helical beams can be switched and their circular-polarization handedness reversed. Meanwhile, the generated vector beams exhibit pronounced azimuthal rotation and polarization evolution during longitudinal propagation, and beams with different helix numbers rotate at different rates. These results verify the feasibility of achieving multi-degree-of-freedom control of terahertz vector beams with a cascaded metasurface, and provide a new phase-design approach for the dynamic reconfiguration of helical vector fields.

**Acknowledgements**: This work is granted by National Natural Science Foundation of China (Grant No. 12404484); Sichuan Province Science and Technology Support Program (Grant Nos. 2025NSFSC2073 and 2025ZNSFSC0846).

**Disclosures**. The authors declare no conflicts of interest.

**Data availability**. Data underlying the results presented in this paper can be obtained from the authors upon reasonable request.

Citation